\documentclass[superscriptaddress,aps,twocolumn,prl,amsmath,amssymb]{revtex4-2}

\usepackage{amsmath}
\usepackage{amssymb}
\usepackage[varg]{txfonts}
\usepackage{color}
\usepackage[breaklinks=true,colorlinks,citecolor=blue,linkcolor=blue,urlcolor=blue]{hyperref}
\usepackage{tikz}
\usetikzlibrary{shapes}
\usepackage{xcolor}

\usepackage{graphicx}
\usepackage{bm}

\makeatletter
\newcommand\bigDiamond{\mathop{\mathpalette\bigDi@mond\relax}}
\newcommand\bigDi@mond[2]{\vcenter{\hbox{\m@th \scalebox{\ifx#1\displaystyle 2\else1.2\fi}{$#1\Diamond$}}}}
\newcommand{\RNum}[1]{\uppercase\expandafter{\romannumeral #1\relax}}

\usepackage{array}
\newcolumntype{M}[1]{>{\arraybackslash}p{#1}}
\newcolumntype{P}[1]{>{\centering\arraybackslash}p{#1}}

\def\XXint#1#2#3{{\setbox0=\hbox{$#1{#2#3}{\int}$}
    \vcenter{\hbox{$#2#3$}}\kern-.5\wd0}}

\def\be{\begin{equation}}
\def\ee{\end{equation}}

\def\bi{\begin{itemize}}
    \def\ei{\end{itemize}}
\def\bn{\begin{enumerate}}
    \def\en{\end{enumerate}}
\def\bea{\begin{eqnarray}}
\def\eea{\end{eqnarray}}
\newcommand{\bpm}{\begin{pmatrix}}
    \newcommand{\epm}{\end{pmatrix}}

\def\ba{\begin{array}}
    \def\ea{\end{array}}
\def\bd{\begin{displaymath}}
\def\ed{\end{displaymath}}

\renewcommand{\imath}{\hspace{1pt}\mathrm{i}\hspace{1pt}}

\begin{document}

\title{Unconventional superconducting pairing in a B20 Kramers Weyl semimetal}

\author{Sougata Mardanya}
\email[Corresponding author:~]{sougata.mardanya@howard.edu}
\affiliation{Department of Physics and Astrophysics, Howard University, Washington, USA}

\author{Mehdi Kargarian}
\email[Corresponding author:~]{kargarian@sharif.edu}
\affiliation{Department of Physics, Sharif University of Technology, Tehran, Iran}

\author{Rahul Verma}
\affiliation{Department of Condensed Matter Physics and Materials Science, Tata Institute of Fundamental Research, Mumbai 400005, India}

\author{Tay-Rong Chang}
\affiliation{Department of Physics, National Cheng Kung University, Tainan 701, Taiwan}
\affiliation{Center for Quantum Frontiers of Research and Technology, Tainan 701, Taiwan}
\affiliation{Physics Division, National Center for Theoretical Sciences, National Taiwan University, Taipei, Taiwan}

\author{Sugata Chowdhury}
\affiliation{Department of Physics and Astrophysics, Howard University, Washington, USA}

\author{Hsin Lin}
\affiliation{Institute of Physics, Academia Sinica, Taipei 11529, Taiwan}

\author{Arun Bansil}
\affiliation{Department of Physics, Northeastern University, Boston, Massachusetts 02115, USA}

\author {Amit Agarwal}
\affiliation{Department of Physics, Indian Institute of Technology, Kanpur 208016, India}

\author{Bahadur Singh}
\email[Corresponding author:~]{bahadur.singh@tifr.res.in}
\affiliation{Department of Condensed Matter Physics and Materials Science, Tata Institute of Fundamental Research, Mumbai 400005, India}

\begin{abstract} 
Topological superconductors present an ideal platform for exploring nontrivial superconductivity and realizing Majorana boundary modes in materials. However,  finding a single-phase topological material with nontrivial superconducting states is a challenge. Here, we predict nontrivial superconductivity in the pristine chiral metal RhGe with a transition temperature of 5.8 K. Chiral symmetries in RhGe enforce multifold Weyl fermions at high-symmetry momentum points and spin-polarized Fermi arc states that span the whole surface Brillouin zone. These bulk and surface chiral states support multiple type-II van Hove singularities that enhance superconductivity in RhGe. Our detailed analysis of superconducting pairing symmetries involving Chiral Fermi pockets in RhGe indicates the presence of nontrivial superconducting pairing. Our study establishes RhGe as a promising candidate material for hosting mixed-parity pairing and topological superconductivity. 
\end{abstract}

\maketitle

The quest for Majorana zero modes (MZMs) which are crucial for the development of fault-tolerant quantum computation devices, has spurred significant interest in the search for topological superconductors~\cite{Nayak2008, Stanescu2013, Elliott2015}. Recent efforts have predominantly focused on characterizing MZMs in theoretical models and on experiments with complex heterostructures~\cite{Read2000, Kitaev2001, Lutchyn2010, Oreg2010}. Unfortunately, the observation of isolated MZMs is still inconclusive. In their simplest form, the MZMs have been predicted to exist at the boundary of a two-dimensional chiral $p-$wave superconductor~\cite{Ivanov2001, Mackenzie2003, Lin2020}. However, the scarcity of such superconductors necessitates the exploration of alternative avenues for realizing MZMs.

A significant breakthrough is the prediction of MZMs at the vortex core of the gapped surface states of three-dimensional (3D) topological insulators under the proximity effect of an $s$-wave superconductor~\cite{Fu2008, Fu2009}. This has ignited efforts to look for superconductivity in topological insulators and semimetals employing chemical doping, external pressure, and reduced dimensionality~\cite{Wray2010, Hor2010, Kirshenbaum2013, Xu2015, Iwaya2017, Roy2019,  Qin2019quasi, Giwa2021, Cook2023, Yang2023}. Noteworthy examples are the iron-based superconductors with Dirac-type inverted band crossings, which show the signature of topological superconductivity with MZMs at the vortex core~\cite{Xu2016, Zhang2018, Kong2019}. Alternately, the momentum-dependent odd-parity pairing can originate from the large spin-orbit coupling (SOC) in non-centrosymmetric superconductors~\cite{Bauer2004, Samokhin2004, Togano2004, Yuan2006, Klimczuk2007, Schnyder2012, Carnicom2018}. However, most superconductors reported to date are inversion symmetric. The Pauli exclusion principle and parity conservation dictate that a superconductor with even parity must be a spin singlet, while an odd parity superconductor must be spin-triplet~\cite{Yip1993, Blount1985}. A material that lacks inversion symmetry and has parity-breaking SOC can thus support an admixture of spin-singlet and spin-triplet paring~\cite{Schnyder2012, Samokhin2004, Togano2004,Yuan2006, Klimczuk2007, Carnicom2018}. 

Motivated by this, we show that the CoSi family of materials are ideal candidates for hosting topological superconductivity. These materials form chiral crystal structures and support unconventional Kramers-Weyl fermions~\cite{chang2018theory, Chang2017, Sanchez2019, Rao2019, Singh2018}. They host the longest possible Fermi-arc surface states spanning the whole surface Brillouin zone (BZ) and feature intriguing optoelectronic and spintronic properties, including helicity-dependent photocurrent~\cite{Rees2020}, a large circular photogalvanic effect~\cite{Ni2021}, a spin Nernst effect~\cite{Hsieh2022}, and exciting plasmons~\cite{PhysRevB.105.165104}. Here, we demonstrate that RhGe exhibits an electron-phonon coupling (EPC) mediated superconductivity driven by multiple type-II van Hove singularities (VHSs) associated with its chiral bands. The superconducting state, nontrivial electronic topology, and chiral crystal symmetries offer all the crucial prerequisites for hosting mixed-parity topological superconductivity in RhGe~\cite{Lee2021, Mandal2023, Yao2015, Qin2019}. By characterizing possible pairing symmetries in Fermi pockets around Kramer Weyl points ($\Gamma$ and $R$ points), we demonstrate a time-reversal symmetry broken topological superconducting state in RhGe.  

\begin{figure}[ht!]
\includegraphics[width=0.99\linewidth]{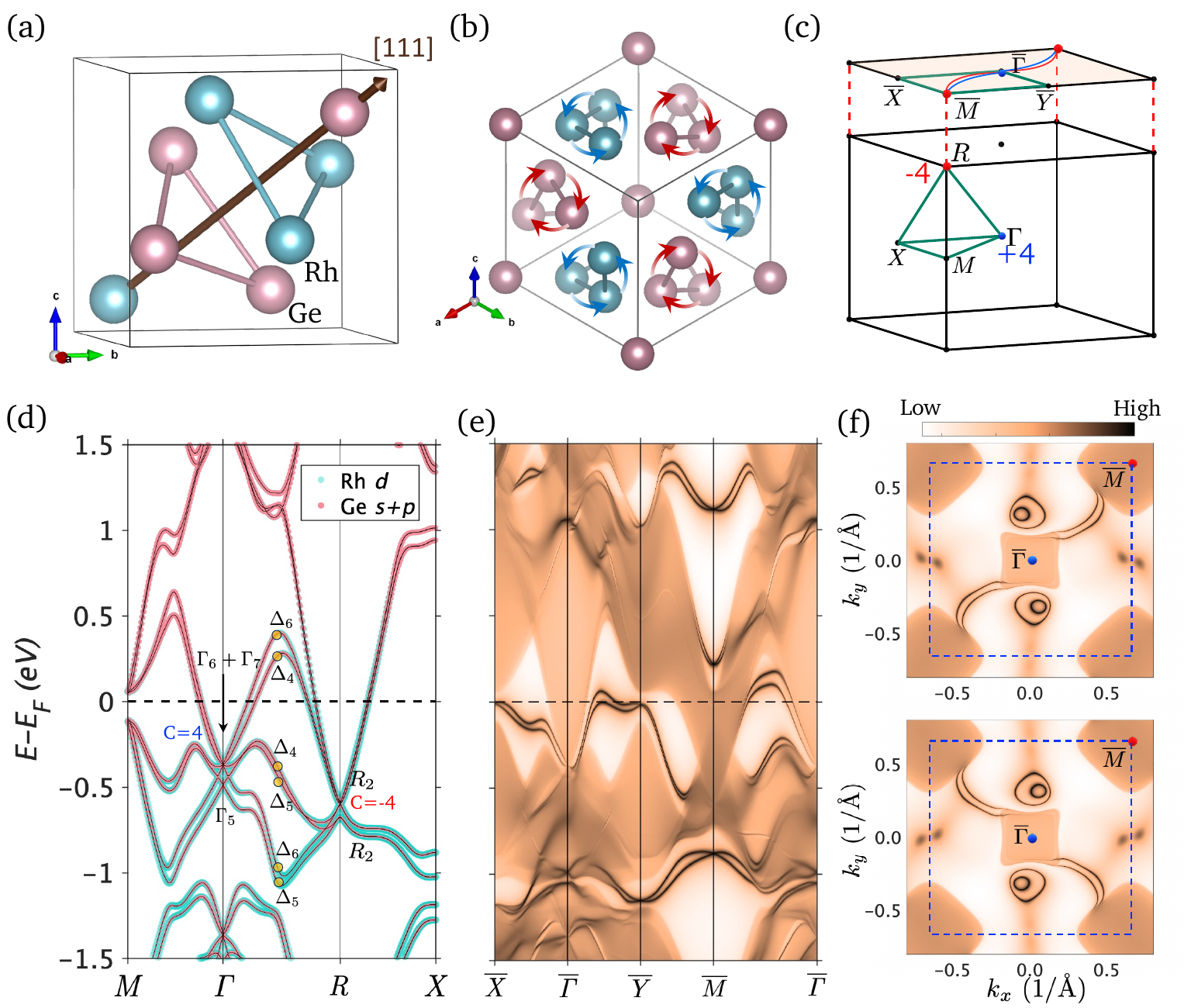} 
\caption{(a)-(b) Chiral crystal structure of RhGe with a pair distortion along [111] chiral axis. (c) The bulk Brillouin zone (BZ) and the projected (001) surface BZ. The high symmetry $k$ points, chiral charges, and the Fermi-arc surface states are highlighted. (d) Bulk band structure of RhGe with spin-orbit coupling. The Rh-$d$ and Ge-($s,p$) orbital contributions and band representations are shown. (e) The (001)-surface spectral function along the high symmetry directions in the surface BZ. Sharp black lines show the chiral surface states. (f) Calculated band countors at the Fermi level. The Fermi-arc surface states connect the +4 chiral charge Weyl node at the BZ center $\overline{\Gamma}$ with the -4 chiral charge node at the BZ corner $\overline{M}$. The top and bottom panels depict the opposite connectivity of Fermi-arc states at the top (Rh termination) and bottom (Ge termination) surfaces.}\label{fig1}
\end{figure}

Beginning with structural chirality and multifold fermions, RhGe forms a B20-type chiral cubic lattice with space group $\mathrm{P2_13}$ (Fig.~\ref{fig1}(a))~\cite{Tsvyashchenko2016, Kamaeva2022}. The B20 structure emerges due to a pair sublattice distortion along the $[111]$ direction (chiral axis), which dimerizes the bonds and breaks (roto)-inversion symmetries [Figs.~\ref{fig1}(a)-(b)]. It preserves a three-fold rotation $\mathrm{\{C_{3, 111}|000\}}$ symmetry along $[111]$ direction and three $\frac{2\pi}{3}$ screw rotation symmetries with a non-primitive translation ($\mathrm{\{S_{2, 100}|\frac{1}{2}\frac{1}{2}0\}}$, $\mathrm{\{S_{2, 010}|0\frac{1}{2}\frac{1}{2}\}}$, and $\mathrm{\{S_{2, 001}|\frac{1}{2}0\frac{1}{2}\}}$). 
The fully relaxed lattice parameter of RhGe is $a=4.79~\mathrm{\AA}$ and internal parameters are $u_{Rh} = 0.134$ and $u_{Ge} = 0.838$ (both Rh and Ge occupy the Wyckoff position $4a \rightarrow \{u_{x},u_{x},u_{x}\}$, $x$ is Rh or Ge), see Supplementary Materials (SMs) for details.\\

The bulk band structure of RhGe is shown in Fig.~\ref{fig1}(d). The presence of time-reversal and non-symmorphic symmetries enforces multiple band degeneracies to generate a unique semi-metallic phase with unconventional multifold fermions. In the absence of SOC, the band structure realizes a three-fold and a four-fold degenerate band-crossing pinned at the high-symmetry $\Gamma$ and $R$ points with chiral charges of +2 and -2, respectively. The inclusion of the SOC splits the three-fold node at $\Gamma$ into a Kramers Weyl node at -0.48 eV and a four-fold degenerate unconventional RSW Weyl node at -0.37 eV. At the $R$ point, a six-fold degenerate double spin-1 chiral node is formed at -0.58 eV. The chiral charges of the nodes at the $\Gamma$  and $R$ points are $+4$ and $-4$, respectively. The surface state spectrum given in Fig.~\ref{fig1}(e) shows four topological surface states connecting the projection of the bulk chiral nodes at the $\overline{\Gamma}$ and $\overline{R}$ points. At the Fermi energy, these states form the Fermi arcs spanning the entire BZ. Unlike the surface states of generic-point Weyl semimetals, the surface states in RhGe originate from the symmetry-enforced bulk Weyl nodes, making them robust against local perturbations and disorder. The chiral nature of these surface states is depicted in the electronic structure of the top and bottom surface terminations, where the Fermi-arc-connectivity is seen to switch between the pockets at the center and corners of the surface BZ [Fig.~\ref{fig1}(f)].

\begin{figure}[ht!]
\includegraphics[width=0.99\linewidth]{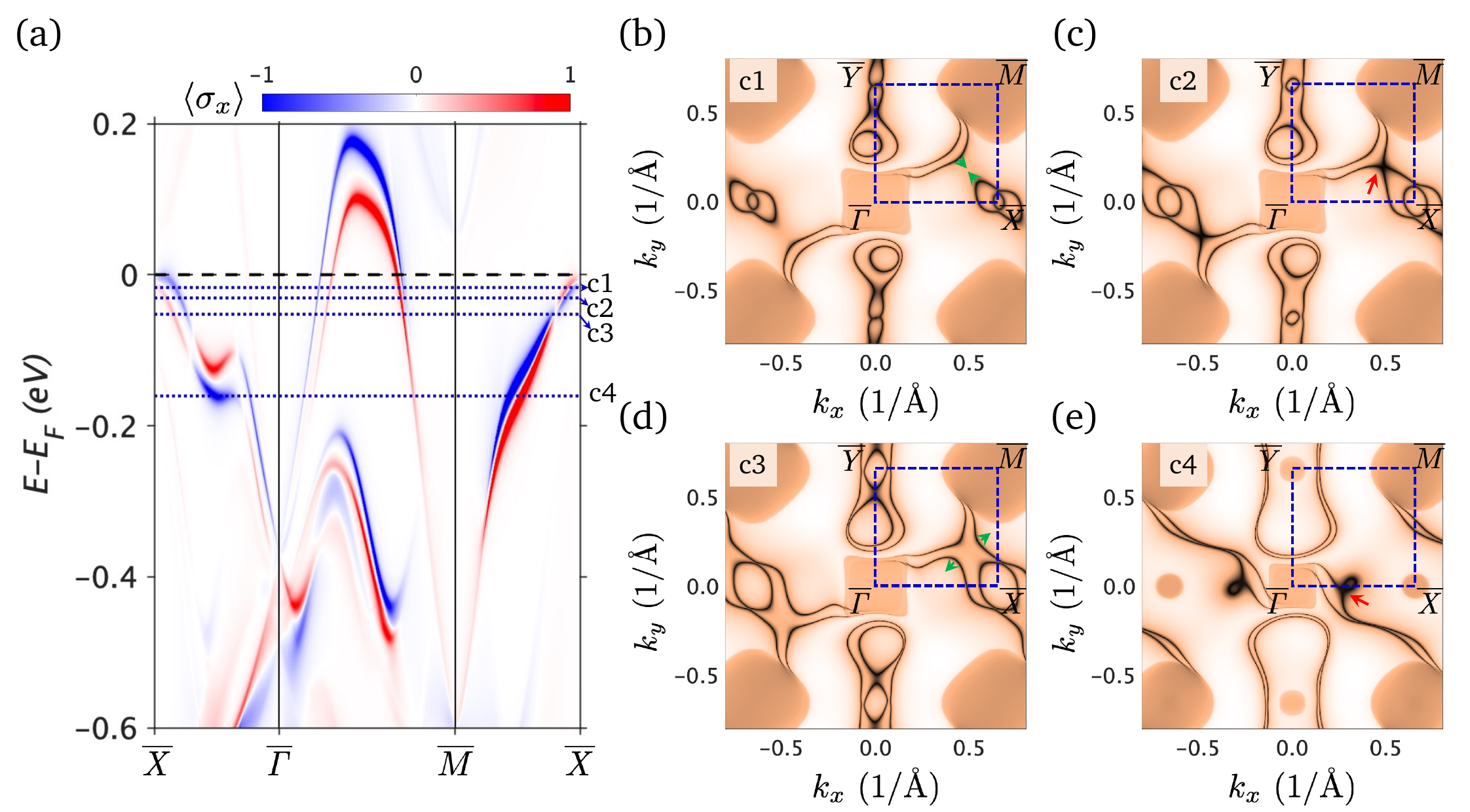} 
\caption{(a) Spin-polarized surface states of RhGe on the (001) surface along $\overline{X}-\overline{\Gamma}-\overline{M}-\overline{X}$.  The dashed blue lines c1 (17 meV), c2 (30 meV), c3 (50 meV), and c4 (160 meV) below the Fermi level show constant energy cuts in (b)-(e). The chiral fermi-arc states exchange connectivity from c1 to c4. Red arrows in c2 and c4 locate the generic $k$-points type II VHSs that appear as a crossing between chiral surface states.}\label{fig2}
\end{figure}

In Fig.~\ref{fig2}, we demonstrate the evolution of surface Fermi arcs as a function of band energy and reveal the presence of type-II VHSs as seen experimentally in CoSi and RhSi~\cite{Sanchez2021}. Note that the position of the VHS at generic $k$ point dictates the nature of the VHS as type II. Figure~\ref{fig2}(a) illustrates the calculated spin-polarized surface states along $\overline{X}-\overline{\Gamma}-\overline{M}-\overline{X}$. A large SOC-driven spin-splitting in the chiral surface states is evident with high spin polarization at the Fermi energy. The evolution of the associated Fermi arcs is shown in Figs.~\ref{fig2}(b)-(e) through successive energy cuts, where the VHSs appear as crossing points of different pockets. In the constant-energy cut c1 ($E =17$ meV), one interband and one intraband crossings appear around the $\overline{X}$  and $\overline{Y}$ points, respectively. As we lower energy to 30 meV (cut c2), the two helicoid arc states move toward each other and touch at the generic momentum points $(\pm 0.49, \pm 0.19)~\mathrm{\AA}^{-1}$ to form type-II VHSs. Further lowering the energy to 50 meV (cut c3) captures the intermediate stage of the chirality exchange between the helicoid states, where the inner state along $\overline{\Gamma}-\overline{Y}$  cross at $(0,\pm 0.52)~\mathrm{\AA}^{-1}$. Finally, at an energy 160 meV (cut c4), another intraband crossing appears at symmetry equivalent generic momentum point $(\pm 0.26, \mp 0.02)~\mathrm{\AA}^{-1}$. These multiple type-II VHSs in the surface state spectrum could induce surface electronic instability and lead to unconventional superconductivity in RhGe.

\begin{figure}[ht!]
\includegraphics[width=0.99\linewidth]{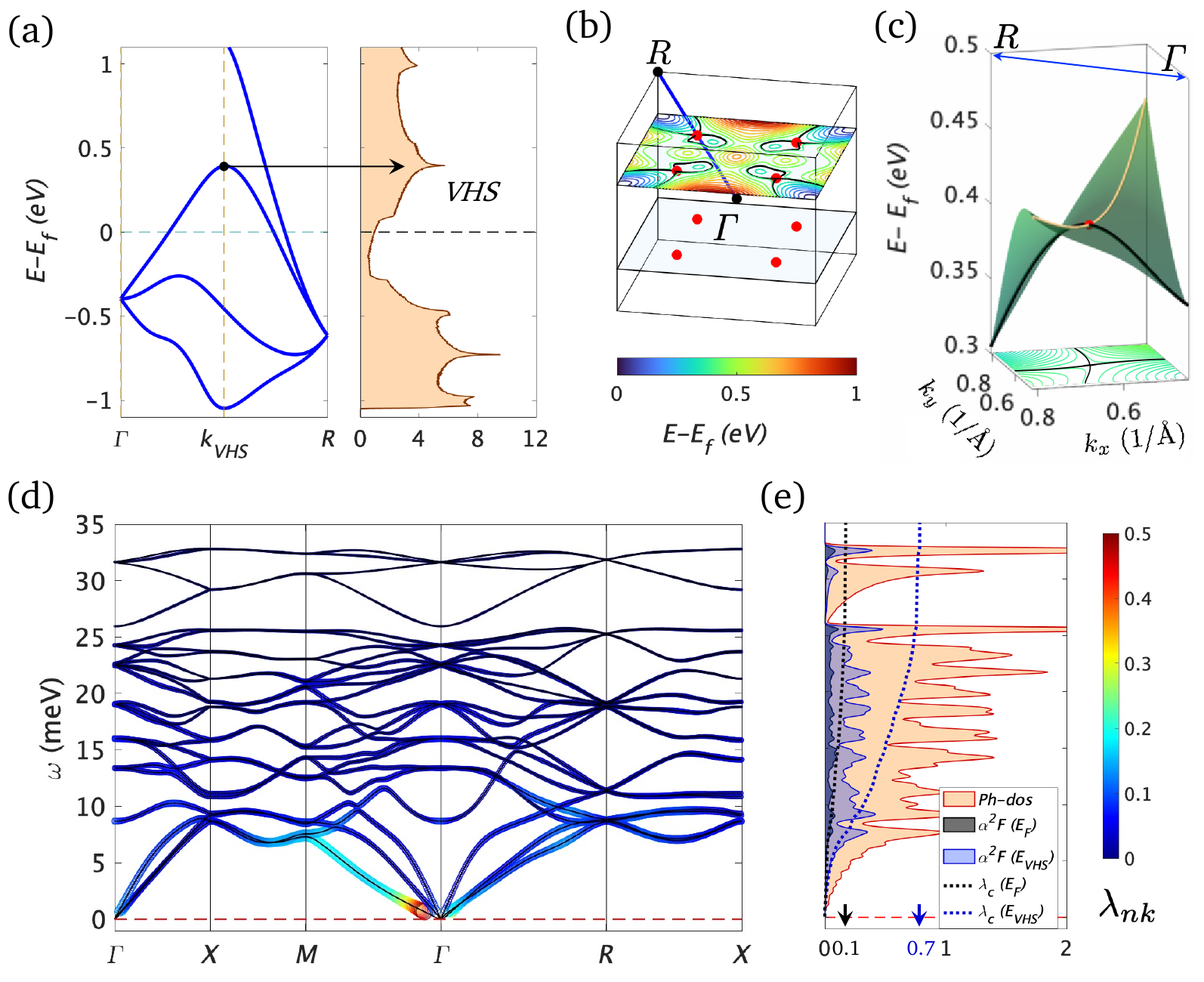} 
\caption{(a) Bulk band structure of RhGe connecting unconventional Weyl fermions at the $\Gamma$ and $R$ points (left panel) and electronic density of states (right panel). Location of bulk VHSs on the $\Gamma-R$ line is marked. (b) The constant energy surface at $k_z=0.62~\mathrm{\AA}$. Locations of VHS in the bulk BZ are marked with red circles. (c) Three-dimensional dispersion around a VHS point on the $k_x-k_y$ plane at $k_z=0.62~\mathrm{\AA}$. (d) Phonon dispersion along the high symmetry directions in the bulk BZ. The size and color of markers indicate band-resolved electron-phonon coupling (EPC) strength. (e) The Eliashberg spectral function and cumulative EPC at the Fermi energy ($E_f$) and the VHS energy ($E_{vhs}=0.4$ eV). The enhancement of electron-phonon interaction due to peaked DOS at VHS is visible in the Eliashberg spectral function and EPC.}\label{fig3}
\end{figure}

Since RhGe exhibits a superconducting state below $\sim4.3$ K~\cite{Tsvyashchenko2016}, its surface states can become superconducting via the proximity effect. Accordingly, we now investigate the possibility of superconductivity in the bulk states and discuss the associated pairing symmetries. Figure~\ref{fig3}(a) shows the bulk band structure along the chiral axis ($\Gamma-R$ line) and density of states (DOS) of RhGe. A careful analysis reveals that a large DOS at 0.4 eV above the Fermi energy is associated with the VHS. In Fig.~\ref{fig3}(b), we illustrate constant energy contours on the $k_z= 0.62~\mathrm{\AA^{-1}}$ plane. The crossing of constant energy lines reveals saddle points at symmetry equivalent generic momentum point $k_{VHS}=(0.62, 0.62, 0.62)~\mathrm{\AA^{-1}}$ on the $\Gamma-R$ line. The associated saddle-like energy dispersion is shown in Fig.~\ref{fig3}(c). Importantly, unlike the usual saddle points, the saddle points here lie on the chiral axis and on the bands forming the multifold Weyl cones. Considering the symmetry of the lattice, there are eight such saddle points in the whole BZ. 

\begin{figure}[ht!]
\includegraphics[width=0.99\linewidth]{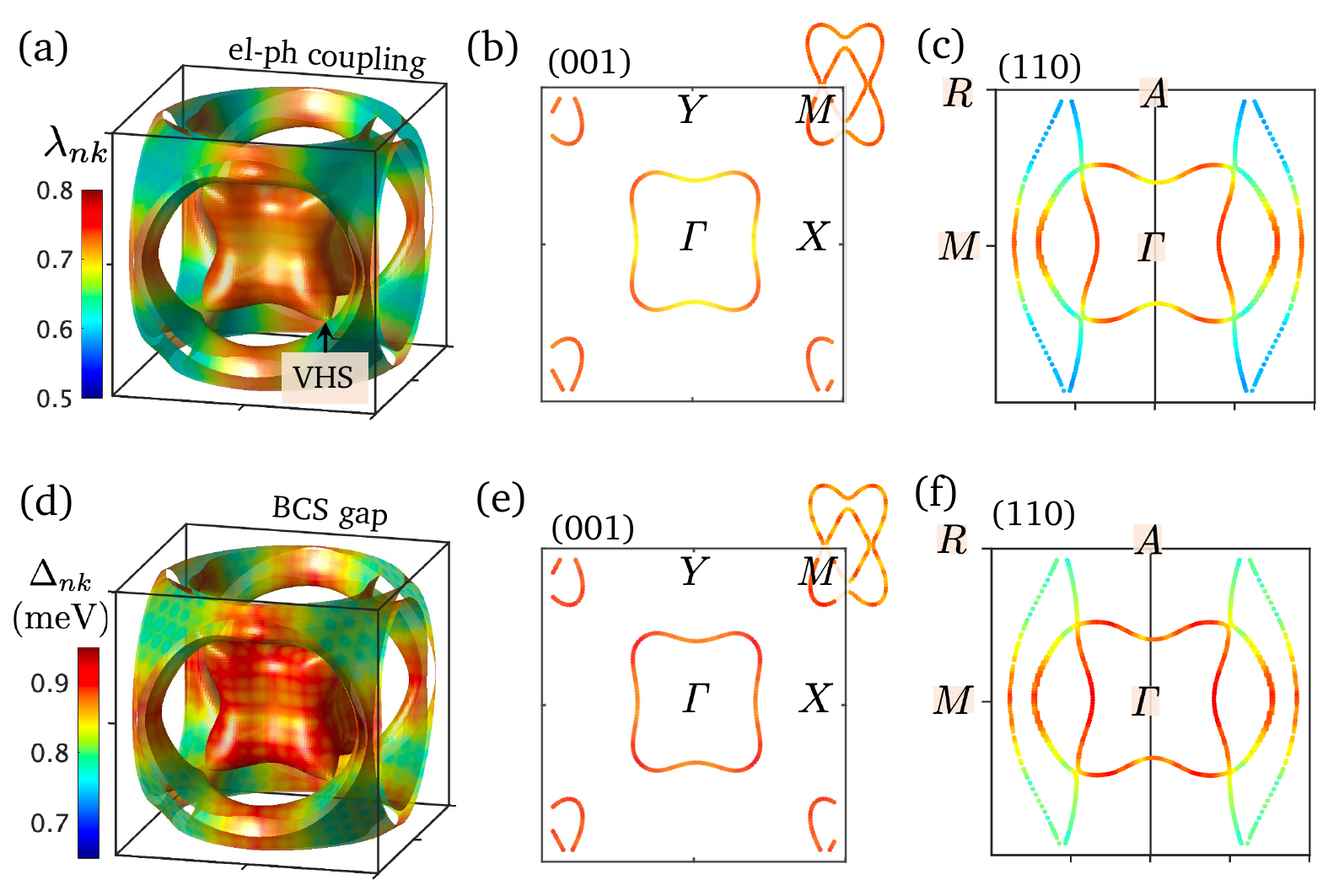} 
\caption{(a) Constant energy surface (CES) of RhGe at $E_{vhs} = 0.4$ eV. The color scale represents the strength of electron-phonon coupling. The VHS located at the [111] chiral axis is marked with a black-red arrow. (b)-(c) The electronic-phonon coupling weighted CES on (001) and (110) planes.  The major contribution to EPC comes from the pockets around $\Gamma$ point. (d) Distribution of superconducting gap over the CES. (e)-(f) The superconducting gap distribution on (001) and (110) plane cuts of the CES. The anisotropic distribution of EPC and the superconducting gap is evident.} \label{fig4}
\end{figure}

The novel electronic phenomena originating from the saddle-point VHSs can be probed by tuning the binding energy at the $E_{VHS} = 0.4$ eV. This can be achieved by electron doping as reported in recent experiments~\cite{Sanchez2021}. To confirm the effects of VHS on the enhancement of electron-phonon interaction, we calculate the phonon spectrum and the momentum resolved EPC $\lambda_{\mathbf{q} \nu}$, where $\mathbf{q}$ is the wavevector, $\nu$ is the phonon band index (see SM for details). The EPC-weighted phonon spectrum of RhGe is shown in Fig.~\ref{fig3}(d), which indicates the large contribution of EPC on an LA branch along the $\Gamma-M$ direction. The phonon DOS, Eliashberg spectral function, and cumulative EPC display an enhanced EPI strength from $\lambda_{E_F}=0.17$ to $\lambda_{E_{VHS}}=0.78$ as one tune energy from Fermi level to VHS. This enhancement of EPC strengthens the BCS superconducting state in RhGe.  An estimation of the superconducting transition temperature using the Allen-Dynes equation, $T_{c}=\frac{\omega_{\log }}{1.2} \exp \left[\frac{-1.04(1+\lambda)}{\lambda\left(1-0.62 \mu^{*}\right)-\mu^{*}}\right],$ with $\lambda_{E_{VHS}}=0.78$ and screened Coulomb interaction parameter $\mu^{*}=0.1$, is found to be $\approx 5.98$ K, close to experimental reported $T_c$~\cite{Tsvyashchenko2016}. Here $\omega_{\log }$ is the logarithmic average of phonon frequency. 

By solving the anisotropic Migdal-Eliashberg equation on the imaginary axis with 0.25 eV cut-off over the Matsubara frequencies, we further obtain the EPC and superconducting gap structure over the constant energy surface (CES) at $E_{VHS}$ (Fig.~\ref{fig4}). The calculated EPC and superconducting gap function over CES are shown in Figs.~\ref{fig4}(a) and \ref{fig4}(c), respectively. Both the EPC and superconducting gap function show anisotropic distribution over the CES in momentum space. This is more evident from the (001) and (110) cuts presented in Figs.~\ref{fig4}(d)-\ref{fig4}(f) and Fig.~\ref{fig4}(i)-\ref{fig4}(k) for EPC and superconducting gap function, respectively. Note that without SOC, EPC is the largest for the pockets centered at the $\Gamma$ point with an anisotropic s-wave-like BCS superconducting gap distribution on the $k_x-k_y$ plane and along the $k_z$ direction. A more realistic description of the gap function over the fermi surface may require calculations with SOC, which is beyond the computational reach of current first-principles methods.

Therefore, to gain a deeper understanding of the gap function and potential pairing symmetries influenced by SOC for the Fermi pockets near the $\Gamma$ point, we employ the kinetic model introduced in Ref.~\cite{Lee2021}, see SM for details. The projected electronic states near the Fermi surface are described by $\hat{\psi}_{m,\mathbf{k}}$ ($m=3/2, 1/2$). These two Fermi pockets are associated with the multifold fermionic excitations described by the total angular momentum $S=3/2$ near the $\Gamma$ point. Ignoring the inter-pocket pairings, the intra-pocket Cooper pairings between the electrons at $\mathbf{k}$ and $-\mathbf{k}$ are described by the condensate $b_{m,\mathbf{k}}=\langle \hat{\psi}_{m,\mathbf{k}}\hat{\tilde{\psi}}_{m,\mathbf{k}}\rangle$. Here, $\hat{\tilde{\psi}}_{m,\mathbf{k}}$ is the time-reversal partner of the electron states at $\mathbf{k}$: $\hat{\tilde{\psi}}_{m,\mathbf{k}}=\eta_{m,\mathbf{k}} \hat{\psi}_{m,-\mathbf{k}}$ with $\eta_{m,\mathbf{k}}=e^{2im\phi_{\mathbf{k}}}$, where $\phi_{\mathbf{k}}$ is the azimutual angle of $\mathbf{k}$. Note that the condensate is an even function of momentum $b_{m,-\mathbf{k}}=b_{m,\mathbf{k}}$ for half-integers $m=3/2, 1/2$, forcing the gap function $\Delta_m(\mathbf{k})=-\sum_{\mathbf{p}} \sum_{m^{\prime}} U_{m m^{\prime}}(\mathbf{k}, \mathbf{p}) b_{m^{\prime}, \mathbf{p}}$ to be even. This can be seen by considering the mean-field theory to decouple the interactions in the Cooper channel as 
$H_{\text {pairing }}=\frac{1}{2} \sum_{\mathbf{k}, \mathbf{p}} \sum_{m, m^{\prime}} U_{m m^{\prime}}(\mathbf{k}, \mathbf{p})$ $\left(\hat{\tilde{\psi}}_{m, \mathbf{k}}^{\dagger} \hat{\psi}_{m, \mathbf{k}}^{\dagger} b_{m^{\prime}, \mathbf{p}}+\hat{\psi}_{m^{\prime}, \mathbf{p}} \hat{\tilde{\psi}}_{m^{\prime}, \mathbf{p}} b_{m, \mathbf{k}}^*-b_{m^{\prime}, \mathbf{p}} b_{m, \mathbf{k}}^*\right)$ with $U_{m m^{\prime}}(\mathbf{k}, \mathbf{p})=U_{m m^{\prime}}(\mathbf{k}, -\mathbf{p})$. The superconducting gap function $\Delta_m(\mathbf{k})=-\sum_{\mathbf{p}} \sum_{m^{\prime}} U_{m m^{\prime}}(\mathbf{k}, \mathbf{p}) b_{m^{\prime}, \mathbf{p}}$ is clearly even $\Delta_m(-\mathbf{k})=\Delta_m(\mathbf{k})$.  
Thus, for superconducting pairing symmetry, $L$ is an even integer which at the lowest angular momentum is $s$-wave ($L=0$) and $d$-wave ($L=2$).  The nature of superconducting gap symmetry is determined by the most attractive channel in expanding the interaction $U_{m m^{\prime}}(\mathbf{k}, \mathbf{p})$ in terms of the basis functions. The $s$-wave pairing is the most symmetric where the gap function preserves the time-reversal symmetry and yields a trivial superconducting state. Higher-angular momentum pairings can break the time-reversal symmetry and give rise to nontrivial superconductivity characterized by the topological charge $\nu$. Different possible pairing symmetries and topological states are summarized in Table~\ref{tab1}. These results indicate that RhGe could realize topological superconductivity. 

\begin{table}[ht!]
\begin{centering}
\caption{Possible superconducting pairing symmetries around $\Gamma$ and the associated topological states.}
\resizebox{\columnwidth}{!}{%
\begin{tabular}{P{1.0cm}P{2.0cm}P{3.0cm}P{2.0cm}P{2.0cm}}
\hline \hline 
$(\mathrm{L}, \mathrm{M})$ & Gap function & Nature of superconducting state & Time-reversal symmetry & Topological charge \\
\hline
$(0,0)$ & $s$-wave & Trivial & Preserved & $\nu=0$ \\
$(2,2)$ & $k_x^2-k_y^2+2 i k_x k_y$ & Weyl  & Broken & $\nu=+2$ \\
$(2,1)$ & $\left(k_x+i k_y\right) k_z$ & Weyl+nodal loop  & Broken & $\nu=+1$ \\
$(2,0)$ & $3 k_z^2-k^2$ & nodal loop  & Preserved & $\nu=0$ \\
$(2,-1)$ & $\left(k_x-i k_y\right) k_z$ & Weyl+nodal loop  & Broken & $\nu=-1$ \\
$(2,-2)$ & $k_x^2-k_y^2-2 i k_x k_y$ & Weyl & Broken & $\nu=-2$ \\
\hline
\end{tabular}
}
 \label{tab1}
 \end{centering}
\end{table}

We emphasize that for the pairing symmetries listed in Table~\ref{tab1}, we employed a rotational symmetric effective model. In an actual crystal, however, the pairing functions follow point-group symmetries. Thus, the effective interaction in the pairing channel should be expanded in terms of the basis functions of the corresponding irreducible representations (Fig.~\ref{fig1}(d)). The point group symmetry at $\Gamma$ is described by the double group $T$. The normal group admits conjugate representations, $E$, with even basis functions $\left(d_{z^2} \equiv 2 z^2-x^2-y^2, d_{x^2-y^2} \equiv x^2-y^2\right)$. Both pairings would yield nodal-line superconductors. Due to the conjugate nature of the representation, a phase-locking between them leads to pairings of types $d_{z^2} \pm i d_{x^2-y^2}$, which breaks the time-reversal symmetry. Since such complex pairings replace the nodal loops with nodes, they are energetically favored over $d_{z^2}$ and $d_{x^2-y^2}$ due to the gain in the condensation energy. 

\begin{table}[ht!]
\begin{centering}
\caption{Possible superconducting pairing symmetries around $R$ and the associated topological states.}
\resizebox{\columnwidth}{!}{%
\begin{tabular}{P{1.0cm}P{2.0cm}P{3.0cm}P{2.0cm}P{2.0cm}}
\hline \hline
$(\mathrm{L}, \mathrm{M})$ & Gap function & Nature of superconducting state & Time-reversal symmetry & Topological charge \\
\hline
$(1,1)$ & $k_x+i k_y$ & Weyl  & Broken &$\nu=+1$ \\
$(1,0)$ & $k_z$ & Nodal loop  & Preserved & $\nu=0$ \\
$(1,-1)$ & $k_x-i k_y$ & Weyl  & Broken & $\nu=-1$ \\
\hline
\end{tabular}
}
 \label{tab2}
\end{centering}
\end{table}

Unlike the $\Gamma$ pocket, the condensate is odd around the $R$ pocket. The excitations around $R$ are described by a multifold fermion with $S=1$, and hence $m=1$ for the states near the Fermi surface. The phase factor $\eta_{m,\mathbf{k}}=\eta_{m,-\mathbf{k}}$ is even and gives rise to an odd condensate $b_{m,\mathbf{k}}=-b_{m,-\mathbf{k}}$. Therefore, only odd angular momenta will survive in the expansion of the pairing interaction $U_{m m^{\prime}}(\mathbf{k}, \mathbf{p})$. The possible pairing symmetries are provided in SMs. The point-group symmetry of $R$ is described by double group $C_3$. The normal group has a one-dimensional representation with the basis $p_z \equiv z$. The corresponding pairing symmetry leads to a nodal-loop superconductor. The symmetry group also admits a conjugate representation with basis functions $\left(p_x \equiv x, p_y \equiv y\right)$. The pairing symmetry could be $p_x \pm i p_y$, $i.e.$, the Weyl superconductors (see Table \ref{tab2}). In recent experiments, RhGe was reported to show superconductivity with the coexistence of weak ferromagnetism~\cite{Tsvyashchenko2016}, which can substantiate the topologically non-trivial nature of the superconductivity.

In summary, we have demonstrated RhGe to be a strong candidate for realizing topological superconductivity. Owing to its structural chirality, RhGe possesses multifold Weyl fermions located at the $\Gamma$ and $R$ momentum points on the chiral axis. The bands connecting these momentum points feature type-II saddle-point VHSs with enhanced EPI and superconductivity. By performing the anisotropic Eliashberg analysis, we demonstrate that both the EPC and the BCS gap exhibit pronounced anisotropy across the Fermi surface. On characterizing the pairing symmetries around the $\Gamma$ and $R$ points using a low-energy model, we establish the nontrivial superconducting state in RhGe. Moreover, the lack of inversion symmetry in RhGe can allow the development of an admixture of pairing symmetries and strengthen the unconventional superconducting state. The presence of time-reversal-symmetry-broken topological superconductivity can be verified via future magnetic optical Kerr and $\mu$-SR experiments. Our study establishes that RhGe is a promising B20 Weyl semimetal for exploring topological superconductivity and Majorana fermions.

{\it Acknowledgements.} 
This work is supported by the Department of Atomic Energy of the Government of India under Project No. 12-R$\&$D-TFR-5.10-0100 and benefited from the computational resources of TIFR Mumbai. H.L. acknowledges the support by the National Science and Technology Council (NSTC) in Taiwan under grant number MOST 111-2112-M-001-057-MY3. The work at Howard University was supported by the US Department of Energy, Office of Science, Basic Energy Sciences Grant No. DE-SC0022216. The work at Northeastern University was supported by the US Department of Energy (DOE), Office of Science, Basic Energy Sciences Grant No. DE-SC0022216 and benefited from Northeastern University’s Advanced Scientific Computation Center and the Discovery Cluster, and the National Energy Research Scientific Computing Center through DOE Grant No. DE-AC02-05CH11231. The authors acknowledge the Texas Advanced Computing Center (TACC) at The University of Texas at Austin for providing HPC resources that have contributed to the research results reported in this paper.

\bibliography{RhGe}

\end{document}